\documentclass[prl,twocolumn,twoside,showpacs,superscriptaddress]{revtex4}

\usepackage{graphicx}
\usepackage{epsfig}
\usepackage{bm,amsmath,amssymb}

\long\def\comment#1{ }
\newcommand{\eqn}[1]{Eq.~\eqref{#1}}

\newcommand{\beq}{\begin{eqnarray}}
\newcommand{\eeq}{\end{eqnarray}}

\newcommand{\rmd}{{\rm d}}
\newcommand{\rme}{{\rm e}}

\newcommand{\del}{\partial}

\newcommand{\order}[1]{\mcal{O}{(#1)}}
\newcommand{\mcal}{\mathcal}

\newcommand{\F}{\mcal{P}}

\newcommand{\abar}{\bar{\alpha}}

\begin{document}

\title{ Medium--induced QCD Cascade: Democratic Branching
and Wave Turbulence}

\author{J.-P. Blaizot, E. Iancu, and Y. Mehtar-Tani}

\affiliation{Institut de Physique Th\'{e}orique (IPhT) Saclay, CNRS/URA2306,
F-91191 Gif-sur-Yvette, France}

\email{jean-paul.blaizot@cea.fr; edmond.iancu@cea.fr; yacine.mehtar-tani@cea.fr}

\begin{abstract}

We study the average properties of the gluon cascade generated by an energetic parton propagating through a quark-gluon plasma.  We focus on the soft, medium-induced emissions  which control the energy transport at large angles with respect to the leading parton. We show that the effect of multiple branchings is important. 
In contrast with what happens in a usual QCD cascade in vacuum, medium-induced branchings are quasidemocratic, with offspring gluons carrying  sizable fractions of the energy of their parent gluon. This results in  an efficient mechanism for the  transport of energy toward the medium, which is akin to wave turbulence with a scaling spectrum $\sim 1/\sqrt{\omega}$. We argue that the turbulent flow may be responsible for the excess energy  carried by very soft quanta, as  revealed by the analysis of the  dijet asymmetry observed in Pb-Pb collisions at the LHC.

\end{abstract}
\maketitle

One important phenomenon  discovered recently in heavy ion experiments at the LHC
is that of  {\em dijet asymmetry},  a 
strong imbalance  between the energies of two back-to-back jets. This asymmetry is commonly attributed to the effect of the interactions of one of the two jets with the hot QCD matter that it traverses, while the other leaves the system unaffected. Originally identified \cite{Aad:2010bu,Chatrchyan:2011sx} as missing 
energy, 
this phenomenon has been subsequently shown
\cite{Chatrchyan:2012ni} to consist in the transport of a 
sizable part of the jet energy by soft particles toward large angles.
Some of the features of in-medium jet propagation are well accounted for by the BDMPSZ  mechanism for medium-induced radiation (from Baier, Dokshitzer, Mueller, Peign\'e, Schiff 
\cite{Baier:1996kr} and Zakharov \cite{Zakharov:1996fv}).
However, most studies within this approach have focused on the energy lost by the
leading particle, while the LHC data call for a more thorough analysis  of  the jet shape for which the  effects of multiple branching at large angles are important.
Within that context, an important step was achieved in  Ref.~\cite{Blaizot:2012fh}, where it was shown that, in  a leading order approximation, one could consider successive gluon emissions as independent of each other. This allows one to treat multiple emissions as a probabilistic 
branching process, in which the BDMPSZ spectrum plays the role of
the elementary branching rate \cite{Baier:2000sb,Baier:2001yt,Jeon:2003gi}.

Specifically, the differential probability per unit time 
and per unit $z$ for a gluon with energy $\omega$ to split into two gluons with 
energy fractions respectively $z$ and  $1-z$ is
 \beq\label{Kdef}
 \frac{\rmd^2 \mcal{P}_{\rm br}}{\rmd z\,\rmd t}\,=\,\frac{\alpha_s}{2\pi}\,
 \frac{P_{g\to g}(z)}{\tau_{_{\rm br}}(z, \omega)},\quad \tau_{_{\rm br}}=\sqrt{\frac{z(1-z)\omega}
 {\hat q_{\text{eff}}}},
 \eeq
where $P_{g\to g}(z)=N_c [1-z(1-z)]^2/z(1-z)$ is the leading order gluon-gluon splitting function,
$N_c$ is the number of colors, $ \hat q_{\text{eff}}\equiv \hat q \left[1-z(1-z)\right]
$, with $\hat q$  the jet quenching parameter (the rate for transverse momentum broadening 
via interactions in the medium), and  $\tau_{_{\rm br}}(z, \omega)$
is the time scale of the branching process. Note that we use 
light-cone (LC)  coordinates and momenta, with the longitudinal axis defined by the direction of motion 
of the leading particle. Correspondingly, the ``energy'' $\omega$ truly refers to the LC 
longitudinal momentum $p^+$ and $t$ to the LC ``time'' $x^+$.  
Equation \eqn{Kdef} applies as long as $\ell\ll \tau_{_{\rm br}}(z, \omega) < L$, where $L$ is the length of the medium, and $\ell$ is
the mean free path between successive collisions.
The second inequality above implies an upper limit on the average energy 
of the offspring gluons: $z(1-z)\omega \lesssim \omega_c$,
where $\omega_c=\hat q L^2/2$ is the maximum energy that can be taken away by a single 
gluon. It follows from Eq.~(\ref{Kdef}) that the probability for having just one emission throughout the medium 
is (for $z$ not too close to $1$) 
$ \sim \abar \sqrt{\omega_c/z\omega}$, 
where $\bar\alpha\equiv \alpha_s N_c/\pi$.
When this becomes of $\order{1}$, i.e, when  $z\omega\lesssim \omega_s\equiv \abar^2\omega_c$, multiple branchings become important.  Note the correlation between the energy $\omega$ of the emitted
gluon and the emission angle $\theta_{_{\rm br}}$ with respect to the jet axis: one has $\theta_{_{\rm br}}
\simeq (2\hat q/\omega^3)^{1/4}$, showing that soft gluons are emitted at large angles.
This correlation will be important for the physical interpretation of our results. 

It will be  useful to express the energy $\omega$  
of a radiated gluon in terms of the energy fraction $x\equiv\omega/E$ of the initial energy $E$
and to replace the light-cone time $t$  by the dimensionless variable  
\beq\label{lambda}
\tau\,\equiv\,\bar \alpha\sqrt{\frac
 {\hat q}{E}}\,t=\bar\alpha\sqrt{2\,x_c}\,\frac{t}{L}\,,\eeq
where $x_c\equiv \omega_c/E$. We restrict ourselves here to the case 
 $E < \omega_c$, i.e., $x_c>1$,  leaving the discussion of the 
$E > \omega_c$ case to a forthcoming publication. Note that the maximal value of $\tau$ is $\tau_{\rm max}=\bar\alpha\sqrt{2\,x_c}$, corresponding to $t=L$.
Then, the branching probability (\ref{Kdef}) can be written as   \beq\label{Klam}
 \frac{\rmd \mcal{P}_{\rm br}}{\rmd z\,\rmd \tau}=\frac{1}{2}\,\frac{\mcal{K}(z)}
 {\sqrt{x}}
 \,,\eeq
where  $\mcal{K}(z)\,\equiv\,{f(z)}/{[z(1-z)]^{3/2}}=\mcal{K}(1-z)$ and $f(z)\equiv\big[ 1-z(1-z)\big]^{5/2}$.
  
In this Letter, we focus on one observable that characterizes the average properties of the in-medium cascade: the  gluon spectrum,  $ D(x,\tau)\equiv x({\rmd N}/{\rmd x})$, with $N$ the number of gluons.  
By exploiting the fact that successive branchings are independent \cite{Blaizot:2012fh} and using standard
techniques for classical branching processes \cite{Cvitanovic:1980ru}, one can show 
that  $D(x,\tau)$ obeys the  following evolution equation 
 \begin{align}\label{eqDf}
   \frac{\del D(x,\tau)}{\del\tau}
  =\int \rmd z \,{\cal K}(z)  
  \left[\sqrt\frac{z}{x}
  D\Big(\frac{x}{z}, \tau\Big)-\frac{z}{\sqrt{x}}\,
  D\big({x},\tau\big)\right].
  \end{align}
  The initial condition corresponds to a single gluon carrying all the energy, that is, 
${D}(x,\tau=0)=\delta(x-1)$. We shall  refer to the right-hand side of Eq.~(\ref{eqDf}) as the ``collision term'' and denote it as ${\cal I}[D]$.
Its physical interpretation  is clear:
The first  contribution,  which is nonlocal in $x$ (except when $x$ is close to 1), is a
{\em gain term}: it describes the rise in the number of gluons
at  $x$ due to emissions from gluons at larger
$x$. Note that the function $D(x,\tau)$ has support only for $0\le x\le 1$,
which limits the first $z$ integral in \eqn{eqDf} to $x< z < 1$.
The second contribution to the collision term,  local
in $x$, represents a {\em loss term}, describing the reduction in the
number of gluons at $x$ due to their decay into gluons with smaller  $x$.
Taken separately, the gain term and the loss term
in \eqn{eqDf} have end point singularities at $z=1$, but these singularities exactly cancel between
the two terms and the overall equation is well defined.

For  $\tau\ll 1$, we may attempt to solve  \eqn{eqDf}  in perturbation
theory, i.e., by iterations. Thus, by substituting, in the collision term, $D(x,\tau)$ by  its  initial value $D^{(0)}(x)=\delta(x-1)$, one obtains  (for $x<1$)
\beq\label{Ds0}
 D^{(1)}(x,\tau)\,=\,\frac{\tau\, f(x)}{\sqrt{x} (1-x)^{3/2}}.
 \eeq
For $\tau=\tau_{\rm max}$,  this is just   the BDMPSZ spectrum. For reasons that will become clear shortly, we refer to the small-$x$ part of this spectrum as the ``scaling spectrum,'' i.e.,  $D_{\rm sc}(x)\sim 1/\sqrt{x} $. {\it A priori}, because one expects the  small-$x$  region of the spectrum to be populated  by multiple branchings, leading to a breakdown of perturbation theory when $x\lesssim \tau^2$, one could
also expect the spectrum to be strongly modified in this region. 
As we shall see,  this is not at all the case: the scaling spectrum remains remarkably stable
(see also Ref. \cite{Baier:2000sb} for a similar observation).

 \begin{figure}[h]
	\centering
	\includegraphics[width=7.5cm]{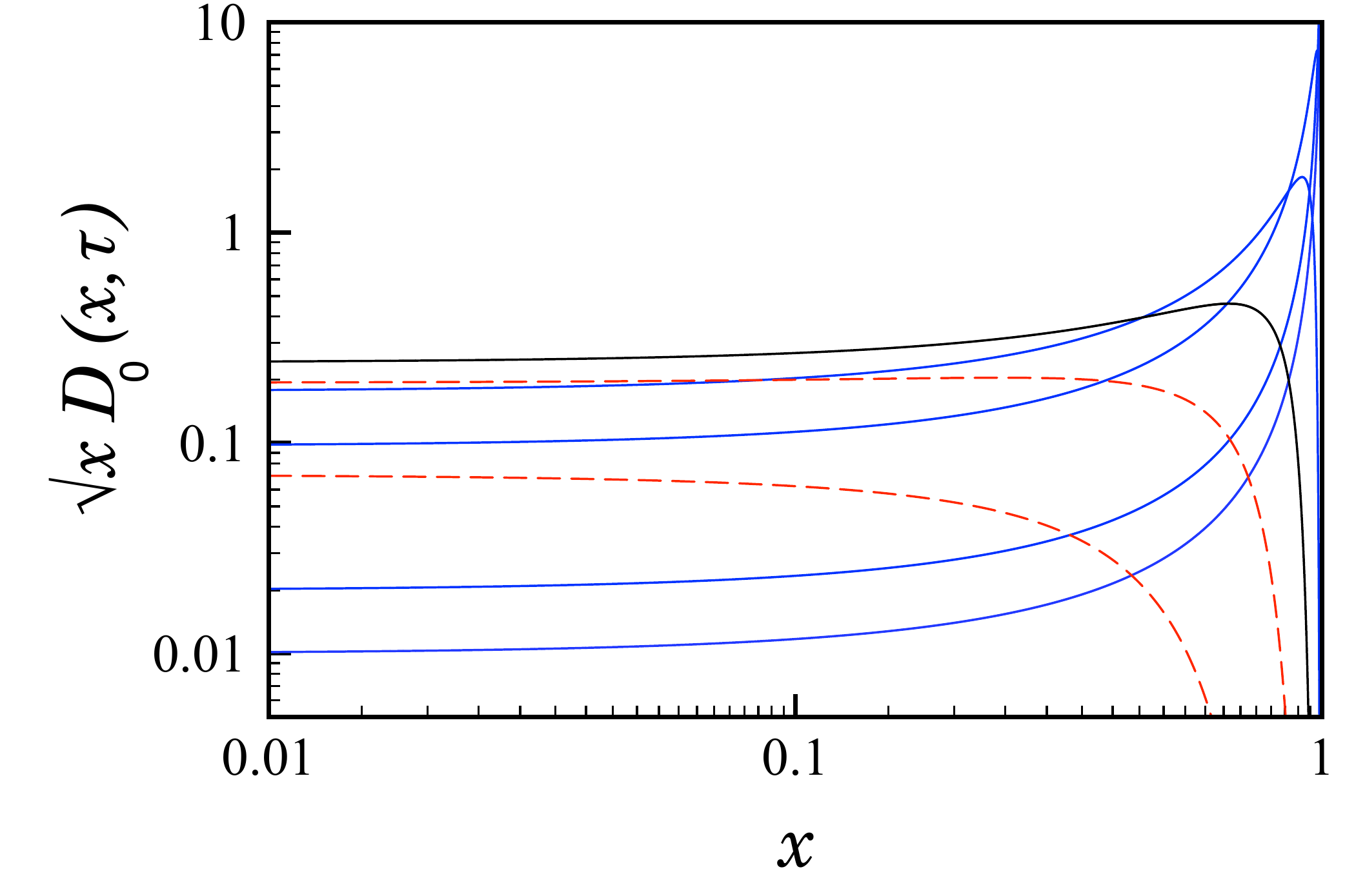}
		\caption{Plot (in log-log scale) of $\sqrt{x} D_0(x,\tau)$, with $D_0(x,\tau)$ given by Eq.~(\ref{Dexact}), as a function of $x$ for various values of $\tau$ (full lines from bottom to top, $\tau=0.01,0.02,0.1,0.2,0.4$; dashed lines from the top down: $\tau=0.6,0.9$).}
		\label{fig1}

\end{figure}

In order to go beyond perturbation theory and  get insight into the nonperturbative features of 
Eq.~\eqref{eqDf},  we have considered a simpler version of this equation, 
obtained by modifying the kernel to ${\cal K}_0(z)= 1/[z(1-z)]^{3/2}$ [i.e, replacing the smooth  function 
$f(z)$ by $1$ in \eqn{Klam}]. This simplification does not affect the singular behavior of the kernel
near $z=0$ and $z=1$, which determines the qualitative features of the solution, but it allows us to solve Eq.~\eqref{eqDf} exactly,  via a Laplace transform. The solution   reads
\beq\label{Dexact}
  D_0(x,\tau)\,=\,\frac{\tau}{\sqrt{x}(1-x)^{3/2}}\ \rme^{-\pi[ \tau^2(1-x)]}\,.\eeq  
 The essential singularity at $x=1$ is a nonperturbative effect that can be understood as a Sudakov suppression factor  \cite{Baier:2001yt}
 (i.e., the vanishing of the probability to emit no gluon in any finite time). 
 Aside from this exponential factor, one recognizes the scaling spectrum which 
$D_0(x,\tau)$  is proportional to at small $x$. This is  illustrated in Fig.~\ref{fig1}: we see that the scaling spectrum is established early on and remains stable as time progresses.  For small times, its amplitude  grows linearly with $\tau$:  the system can then be viewed as a radiating source located at $x\lesssim 1$ and feeding all the small-$x$ modes. As time passes, the source weakens and eventually disappears into the left moving ``shock wave'' visible  in Fig.~\ref{fig1}. 

Another important feature of the branching dynamics illustrated in Fig.~\ref{fig1} is the fact that
the total energy which is stored in the spectrum 
(i.e., in the gluon modes with  $0<x<1$) decreases with time:
${\cal E}_0(\tau)\equiv \int_0^1 \rmd x D_0(x,\tau)={\rm e}^{-\pi\tau^2}$.  This is related to the existence of a 
scaling solution, as alluded to above: the fact that the spectrum keeps the same shape at small $x$
when increasing $\tau$ implies that the energy flows from higher to lower values of $x$ without
accumulating  at any value $x>0$. This should be contrasted to what happens in standard parton
cascades,  like that  described by the DGLAP equation 
\footnote{The
DGLAP equation \cite{DGLAP} can be recovered from
\eqn{eqDf} by substituting ${\cal K}(z)/\sqrt{x}\to 1/[z(1-z)]$ and interpreting the time as
$\tau=\abar\ln(Q^2_0/Q^2)$, with $Q^2$ the parton virtuality which decreases along the cascade.}. 
In that case, the
spectrum becomes steeper and steeper at small $x$ with increasing evolution time,
and the energy sum rule
$\int_0^1 \rmd x D(x,\tau)=1$ is satisfied at any $\tau$ --- ``the energy remains in the spectrum''.
Returning to the medium-induced branching process, we note that the energy is conserved in that case, too, 
since it is so at each elementary branching. Formally, what happens is that  a ``condensate''  
develops at $x=0$, playing the role of a sink where the excess energy coming 
from the large-$x$ region gets stored. With increasing time, a substantial fraction of the total energy
can thus flow ``outside the spectrum.''

 \begin{figure}[h]

		\includegraphics[width=7.5cm]{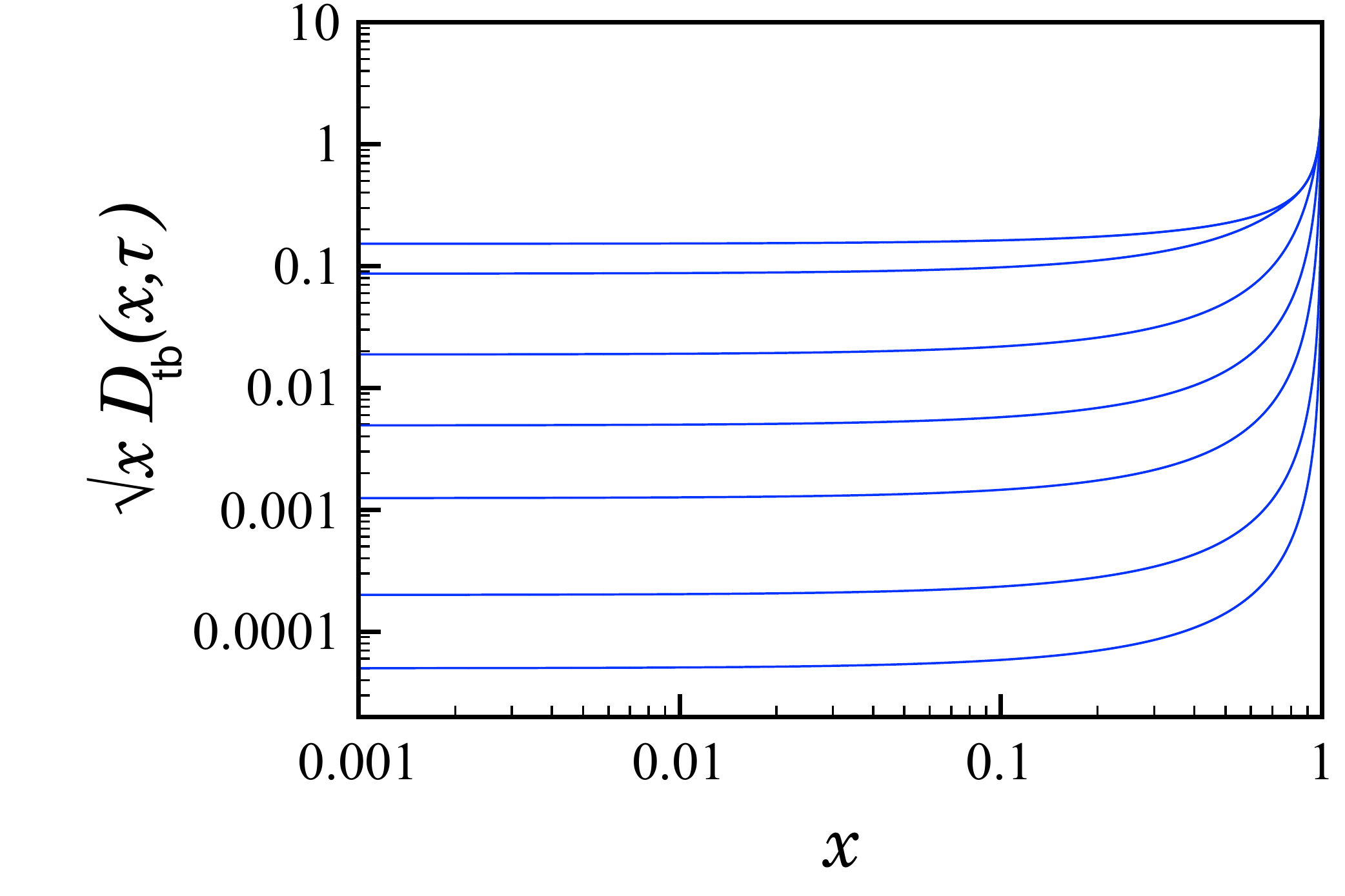} 
		\caption{The function $\sqrt{x} D_{\rm tb}(x,\tau)$ [Eq. (\ref{Dtbex}) for $A=1$]  at
		various times --- from early time, where it resembles Fig.~\ref{fig1}, untill late time, when it approaches the steady state and saturates at the value $A/2\pi$ at small $x$. 
		The values of $\tau$ are, from bottom to top,  0.01, 0.02, 0.05, 0.1, 0.2, 0.5 and 1. }
		\label{fig2}

\end{figure}

To shed more light on this flow phenomenon,
it is instructive to analyze an auxiliary problem --- that of a system driven by a permanent 
source of energy localized at $x=1$. Consider then the equation 
\beq\label{eqtb}
\frac{\del D(x,\tau)}{\del\tau}= A\delta (1-x)+{\cal I}[D],
\eeq
For the simplified kernel ${\cal K}_0$, one readily verifies  that the ``turbulent spectrum'' (see below)
  \beq\label{Dtbex}
  D_{\rm tb}(x,\tau)\,=\,\frac{A}{2\pi \sqrt{x(1-x)}}\ 
  \left(1-\rme^{-\pi[\tau^2(1-x)]}\right)\eeq
solves this equation with initial condition $ D_{\rm tb}(x,\tau=0)=0$ [observe
that the derivative of $D_{\rm tb}(x,\tau)$ is equal to $D_0(x,\tau)$, 
to within the multiplicative constant $A$].
By comparing Figs.~\ref{fig1} and \ref{fig2}, one sees that the behaviors of $D_0(x,\tau)$ 
and $D_{\rm tb}(x,\tau)$ are remarkably similar at small $\tau$. 
However, the most remarkable property of $D_{\rm tb}(x,\tau)$ is 
that it converges to a steady
function, $D_{\rm st}(x)=(A/2\pi) /\sqrt{x(1-x)}$. To understand this, we note that 
$D_{\rm st}(x)$ annihilates (exactly) the collision term, i.e., ${\cal I}[D_{\rm st}]=0$, 
as can be verified by an explicit calculation.  
As time goes on, the solution $ D_{\rm tb}(x,\tau)$ is gradually driven to $D_{\rm st}(x)$, but since 
this fixed-point solution reduces to the scaling spectrum at $x\ll 1$, one observes no change in the
shape at small $x$, but just an overall time-dependent scaling. 

We complete our analysis by calculating  the flow of energy that gets transmitted per unit time from the region $x>x_0$ to the region $x < x_0$. If we denote by ${\cal E}(x_0,\tau)=\int_{x_0}^1\rmd x D(x,\tau)$ the total energy that is contained in the modes with $x>x_0$ and recognize that the rate of change of ${\cal E}(x_0,\tau)$ is due both to a possible  source of strength $A$ localized at $x=1$ 
and to the flow $ \F(x_0,\tau)$ at $x_0$, we get  the general expression
 \beq\label{Pdef}
  \F(x_0,\tau)&\equiv&A-\frac{\del {\cal E}(x_0,\tau)}{\del\tau}=- \int_{x_0}^{1}\rmd x\,{\cal I}[D].\eeq
 An explicit calculation for $D=D_{\rm tb}$ yields 
 \beq\label{Pdef2}
 \F(x_0,\tau)=A\left[1-{\rm e}^{-\pi \tau^2} {\rm erfc}\left(\sqrt{\frac{\pi x_0}{1-x_0} }\tau  \right)     \right],
 \eeq
 where ${\rm erfc}(x)$ denotes the complementary error function. In order to analyze the physical content of this expression, it is actually useful to rewrite the integral of the collision term in Eq.~(\ref{Pdef}) in the following form :
\begin{align}\label{Px0}
  \F(x_0,\tau)=\int_0^1  \rmd z\,z\,{\cal K}(z)\int_{x_0}^{{\rm min}(1,x_0/z)} \rmd x\,
  \frac{D\left(x,\tau\right)}{ \sqrt{x}}.
 \end{align}

At small times, $\pi\tau^2\ll 1$, and for $x_0$ not too close to either $0$ or $1$, one can use the expansion ${\rm erfc}(x)\simeq 1-2x/\sqrt{\pi}$ in Eq.~(\ref{Pdef2}) and get  
$ \F(x_0,\tau)\simeq 2A \tau\,\sqrt{x_0/(1-x_0)}$.
This result can also be obtained from  Eq.~(\ref{Px0}) by substituting $D(x,\tau)$ with
$D^{(1)}(x,\tau)= A\tau\delta(1-x)$, as appropriate at small time.
Thus, we can interpret this early-time contribution to $\F(x_0,\tau)$ as due to direct radiation 
from the source at $x=1$ toward the various modes at $x_0 < 1$. Note that this
involves branchings with $z\leq x_0$ which, for $x_0\ll1$, are strongly asymmetric ($z\ll 1$).

As time goes on, however,  the distribution of energy among the various modes is such that gain and loss terms equilibrate locally, at which point a steady state is reached with all the energy provided by the source  flowing throughout the entire system and leaving the population of the various modes unchanged. In the steady
regime reached for $\tau\gtrsim 1/\sqrt{\pi}$, the energy flux $\F(x_0,\tau)$ is both stationary
($\tau$ independent) and uniform ($x_0$ independent), and equal to $A$ (the flux inserted by the
source). Actually, this uniform component of the flow develops already at earlier times. It can be
obtained by evaluating Eq.~(\ref{Pdef2}) at $x_0=0$ 
and reads $\F(x_0=0,\tau)=A(1-{\rm e}^{-\pi\tau^2})$. This result can be recovered  
from Eq.~(\ref{Px0}) with $x_0\ll 1$ by approximating $D(x,\tau)$ with the scaling part 
of the spectrum (\ref{Dtbex}), $D(x,\tau)\simeq A(1-{\rm e}^{-\pi\tau^2})/(2\pi\sqrt{x})$, and noting that 
 \beq\label{v}
\upsilon_0\equiv \int_{ 0}^1 \rmd z \frac{1}{\sqrt{z} (1-z)^{3/2}}\ln\frac{1}{z}=2\pi.
 \eeq
  What this second calculation demonstrates is that,   in contrast to what happens for the direct radiation, here the typical branchings  involve the whole range of $z$ values [about half of the value of the integral (\ref{v}) comes from  the range $0.15\lesssim z\lesssim 0.85$]. We refer to this property as ``quasidemocratic branching.''

The properties that we have just discussed, namely, the existence of a steady scaling solution when the system is coupled to a source and, related to it, the presence of a component of the flow that is independent of the energy,
are distinctive signatures of what is known as {\em (weak) wave turbulence} \cite{KST}.  A crucial
ingredient of this phenomenon is the locality of the interactions in momentum space, a property
which in the present case is only marginally satisfied, as quasidemocratic branching. 

We wish to stress that democratic branching is not common 
in standard parton cascades, like the one described by the DGLAP equation, 
which are rather controlled by very asymmetric branchings (with $z$ near 0 or 1).  In particular, one can 
verify that for the DGLAP cascade, the energy flow vanishes when $x_0\to 0$ [roughly like 
$\F(x_0) \sim x_0\ln(1/x_0)$]:  the total energy of the 
cascade remains in the spectrum, as already mentioned. Furthermore, the total energy carried by the 
soft modes at $x \le x_0$ with $x_0\ll 1$ is relatively small, as can be inferred from phase-space considerations. This is very different from the turbulent
cascade studied here, in which a significant fraction of the total energy is transported
below any given value $x_0 >0$, meaning at very large angles.

Many of the features that we have  uncovered by studying the source problem and for
the simplified kernel remain valid without the source, and for the general kernel, as we have verified via an explicit numerical solution. 
We return now to this initial setting and limit ourselves to small times, for which we can obtain analytical estimates. 
The flow, calculated from  Eq.~(\ref{Px0}),  takes the form 
 \beq\label{flowradiat}
  \F(x_0,\tau)\simeq 2\sqrt{x_0}+  \upsilon\tau\,,\eeq
where $\upsilon=4.96$ is given by an integral similar to  that in
\eqn{v} but with the full $f(z)$ in the integrand. One recognizes in \eqn{flowradiat}
the two components that we discussed earlier, that is, the direct radiation ($2\sqrt{x_0}$) and the turbulent flow ($v\tau$). Although formally subleading at small times, 
 the turbulent flow dominates over direct radiation  when 
$x_0\lesssim \tau^2$, that is, in the  region where multiple branchings 
are known to be important. 

The total energy transported by the turbulent flow
can be estimated by integrating  the second term 
of Eq.~(\ref{flowradiat}) over time. Returning to physical units, one gets
\beq
{\cal E}_{\rm flow}=E\frac{\upsilon\tau_{\rm max}^2}{2}=\upsilon \,\abar^2 \omega_c\,,
\eeq
a result which, remarkably, is independent of the energy $E$ of the leading particle.
This turbulent flow is a part of the jet energy that  is not carried by the particles present in the spectrum. It corresponds to what we identified earlier as the energy stored in a condensate at $x=0$. In more
physical terms, we may associate this energy with that transferred to the
medium in the form of very soft quanta emitted at large angles.

It is beyond the scope of this Letter to present a detailed comparison with the data. 
However, the following order-of-magnitude 
estimates should confirm  the relevance of the present discussion for the dijet asymmetry observed at the LHC.  Using the conservative estimate $ \omega_c=40$~GeV
(corresponding to $\hat q=1\,{\rm GeV}^2/{\rm fm}$ and $L\simeq 4$~fm),
together with  $\abar^2\simeq 0.1$, one finds ${\cal E}_{\rm flow}\simeq 20$~GeV, a value that 
compares well with the observations. Indeed, the detailed analysis by CMS 
\cite{Chatrchyan:2012ni} shows that the energy
imbalance between the leading and the subleading jets is compensated by an excess
of semihard ($p_T < 8$~GeV) quanta propagating at large angles, outside the 
cone defining the subleading jet. 
For the most asymmetric events, the total energy in excess is about 25~GeV. Remarkably, most
of this energy (about 80\%) is carried by very soft quanta with $p_T\le 2$~GeV \footnote{This appears to be compatible with a recent Monte Carlo analysis \cite{Apolinario:2012cg}, where multiple gluon branchings are phenomenologically implemented.}. 
This observation would be difficult to reconcile with the hypothesis that these
particles come from gluons in a BDMPSZ-like spectrum (which would imply that most of 
the excess energy would be carried by the hardest gluons with energies $\lesssim 8 $ GeV).
But it could be naturally explained by associating these soft particles with those transported by the turbulent flow that we have discussed in this Letter.

\noindent{\bf Acknowledgements}

We thank A. H. Mueller for useful discussions and correspondence. This research is supported by the European Research Council under the Advanced Investigator Grant No. ERC-AD-267258.

\providecommand{\href}[2]{#2}\begingroup\raggedright
\endgroup

\end{document}